\documentclass[review,authoryear,12pt]{elsarticle}

\usepackage{graphics}

\usepackage{amssymb}

 \usepackage{lineno}
 
 \usepackage{multirow}
 \usepackage{booktabs}
 \usepackage{caption}

\journal{arXiv}

\begin{document}

\begin{frontmatter}



\title{Assessing Bone Quality of Spine on Children with Scoliosis Using Ultrasound Reflection FAI Method {\textendash} A Preliminary Study}






\author[affcec0f472d9904cbbab3f10dd407c4f2e,affcb2e2e0175104cb2a6b6c0dceb9d7f70,aff3c7698ab39874202b7f6adbff36d12c3]{Sheng Song}
\author[affcec0f472d9904cbbab3f10dd407c4f2e,affcb2e2e0175104cb2a6b6c0dceb9d7f70,aff3c7698ab39874202b7f6adbff36d12c3]{Hongbo Chen}
\author[shanghaitech_spst]{Conger Li}
\author[affcef7c2ca9f8940a0b85960ea7af00d23]{Edmond Lou}
\author[afff1a67401b10542279b7df7d3544ff302]{Lawrence H Le}
\author[affcec0f472d9904cbbab3f10dd407c4f2e,aff45e3de57b1e545e98ad7d125a143b7d8]{Rui Zheng\corref{contrib-58ee982724894b80bb276e353ed6f040}}
\ead{zhengrui@shanghaitech.edu.cn}\cortext[contrib-58ee982724894b80bb276e353ed6f040]{Corresponding author.}
    
\address[affcec0f472d9904cbbab3f10dd407c4f2e]{School of Information Science and Technology\unskip, 
    ShanghaiTech University\unskip, 201210\unskip, Shanghai\unskip, China}
  	
\address[affcb2e2e0175104cb2a6b6c0dceb9d7f70]{Shanghai Institute of Microsystem and Information Technology\unskip, 
    Chinese Academy of Sciences\unskip, 200050\unskip, Shanghai\unskip, China}
  	
\address[aff3c7698ab39874202b7f6adbff36d12c3]{
    University of Chinese Academy of Sciences\unskip, 100049\unskip, Beijing\unskip, China}

\address[shanghaitech_spst]{School of Physical Science and Technology\unskip, ShanghaiTech University\unskip, 201210\unskip, Shanghai\unskip, China}
  	
\address[affcef7c2ca9f8940a0b85960ea7af00d23]{Department of Electrical and Computer Engineering\unskip, 
    University of Alberta\unskip, Edmonton\unskip, Canada}
  	
\address[afff1a67401b10542279b7df7d3544ff302]{Department of Radiology and Diagnostic Imaging\unskip, 
    University of Alberta\unskip, Edmonton\unskip, Canada}

\address[aff45e3de57b1e545e98ad7d125a143b7d8]{Shanghai Engineering Research Center of Intelligent Vision and Imaging\unskip, 
ShanghaiTech University\unskip, 201210\unskip, Shanghai\unskip, China}



\begin{abstract}
	Osteopenia is indicated as a common phenomenon in patients who have scoliosis. Quantitative ultrasound (QUS) has been used to assess skeletal status for decades, and recently ultrasound imaging using reflection signals from vertebra were as well applied to measure spinal curvatures on children with scoliosis. The objectives of this study are to develop a new method which can robustly extract a parameter from ultrasound spinal data for estimating bone quality of scoliotic patients and to investigate the potential for the parameter on predicting curve progression. The frequency amplitude index (FAI) was calculated based on the spectrum of the original radio frequency (RF) signals reflected from the tissue-vertebra interface. The correlation between FAI and reflection coefficient was validated using decalcified bovine bone samples in vitro, and the FAIs of scoliotic subjects were investigated in vivo referring to BMI, Cobb angles and curve progression status. The results showed that the intra-rater measures were highly reliable between different trials (ICC=0.997). The FAI value was strongly correlated to the reflection coefficient of bone tissue (R\ensuremath{^{2}}=0.824), and the lower FAI indicated the higher risk of curve progression for the non-mild cases. This preliminary study reported that the FAI method can provide a feasible and promising approach to assess bone quality and monitor curve progression of the patients who have AIS.
\end{abstract}

\begin{keyword} 
Frequency amplitude index\sep Bone quality\sep Reflection coefficient\sep Ultrasound spine imaging\sep Scoliosis
\end{keyword}

\end{frontmatter}

\pagebreak









\section*{Introduction}
\label{intro}
Scoliosis is a three dimensional (3D) spine deformity characterized with lateral curvature and axial vertebral rotation (AVR). Eighty percent of scoliotic cases are idiopathic, and it is usually detected on the juvenile and adolescent stages (JIS or AIS) with the prevalence ranged from 0.5\% to 4\% \citep{cheng2015adolescent,hresko_idiopathic_2013,altaf2013adolescent}. During the rapid growth period of children, 15{\textendash}30\% scoliotic curves progress, therefore it is a major concern for orthopedic surgeons to identify the spine curves with a high risk of progression which require active treatment \citep{soucacos_assessment_1998,van2007scoliosis,kim_scoliosis_2010}. The standing posteroanterior (PA) radiograph has been the standard routine method to diagnose and monitor spine curves for decades. The follow-up radiographs are prescribed every 4{\textendash}12 months depending on the age and growth rate of the scoliotic cases \citep{van2007scoliosis,cobb_outline_1948}.

Many Studies have indicated that lower bone mineral density (BMD) is a common phenomenon in scoliotic patients, especially at the region of spine, femoral neck and calcaneus, 
however no direct correlation has been revealed between the BMD and the scoliotic features, such as curve pattern or curve severity \citep{cook_trabecular_1987,cheng_osteopenia_2001,li_persistent_2020}. 
Recently it has been reported that BMD is an important risk factor affecting curve progression. In the study involving 324 adolescent girls, 50\% subjects showed progressed curves, and osteopenia 
in the femoral neck of the hip was indicated as a factor in the prediction of curve progression with the odd ratio of 2.5 \citep{hung_osteopenia_2005}. Although the dual-energy X-ray absorptiometry (DXA) is the most commonly used method 
to evaluate BMD, it exposes the patients to more ionizing radiation besides the follow-up radiographs, consequently it is unusual to measure BMD in children with scoliosis. On the other hand,
 BMD is influenced by the absorption of bone tissue which is only related to the density information; it cannot report the elastic strength information which reflects the mechanical properties of bone.
  Therefore the parameter is insensitive to detect the change of bone quality and mechanical properties. The change of bone density can be detected using radiography only when over 20-40\% loss of skeletal calcium occurs \citep{castriota-scanderbeg_abnormal_2005}.

  Quantitative ultrasound (QUS) has been deployed to evaluate bone quality especially for the assessment of osteoporosis and prediction of the fracture risk since 1990s \citep{langton_clinical_1996,murashima2020anisotropic,liu2020ultrasonic,grimal_quantitative_2019}. As the two most important ultrasonic evaluation 
  parameters in QUS, broadband ultrasound attenuation (BUA) and speed of sound (SOS) are related to both bone density and mechanical properties. \citet{lam_abnormal_2011,lam_quantitative_2013} calculated the stiffness index (SI) from the measured BUA and SOS on the 
  calcaneus to evaluate bone quality and to predict curve progression on AIS subjects. Among 294 recruited girls followed-up beyond skeletal maturity, the adjusted odds ratio of 
  curve progression for low SI was 2.0, and it proposed that SI can be used as an independent prognostic factor for curve progression instead of BMD \citep{lam_quantitative_2013}. Different from the ultrasound transmission modality applied on the 
  calcaneus, using the spectral amplitudes of the reflection signals from the interface between cortex and marrow, the BUA of cortical bones can be also estimated by the spectral ratio method and peak frequency shift method \citep{zheng_spectral_2007,zheng_broadband_2009}.

  Recently, a new ultrasound (US) imaging method has been developed to measure spine curves on children with scoliosis, and the results showed significant consistency with the measurements from the conventional radiographic 
  method \citep{zheng_improvement_2016,zheng_assessment_2017,he2017effective,ungi2020automatic}. A series of B-scan transvers images were continuously acquired along the subject’s back and then reconstructed to a 3D data volume of spine. Based on the strong reflection from the interface between vertebra 
  and soft tissue, each lamina  was identified and used as the landmarks to assess the spinal curvatures using the center of lamina (COL) method \citep{zheng_intra-_2015}. On the other hand, the amplitude information can be simultaneously collected for the analysis of bone properties and quality. \citet{zheng_estimation_2015} implemented a pilot study using the ultrasound reflection signals from 
  vertebra to assess the bone quality of spine. It exported the amplitude information from the original radio frequency (RF) ultrasound data and calculated the reflection index (RI) of vertebra on 18 subjects with AIS. The RI 
  values of the non-mild curve and mild curve groups were 0.47 and 0.74, respectively.

   The objectives of this study are to provide a more reliable and robust estimation parameter of bone quality derived from the reflection spinal signals, and to investigate how the defined parameter is related to curve severity and curve progression of subjects with scoliosis.
\section*{Materials and Methods}
\label{MaM}
      
\subsection*{Frequency amplitude index (FAI)}
For the ultrasound signals reflected from the interface between soft tissue and cortex of a vertebra,  the reflected wave in the frequency domain \(A\left( f,z \right)\) can be described as a superposition of plane waves \citep{zheng_spectral_2007}:
\begin{equation}
	A(f, z)=R \cdot S(f) \cdot e^{ -\alpha  z} \cdot e^ {i\left(k_{r}  z-2 \pi f t\right)}
	\label{eq1}
\end{equation}
where \(S(f)\) is the source pulse emitted from transducer probe, \(f\) is the frequency of ultrasound wave, \(R\) is the reflection coefficient between soft tissue and vertebra cortex, \(\alpha\) is the attenuation coefficient of soft tissue, 
\begin{figure}
	\centering
	\includegraphics[scale=0.6]{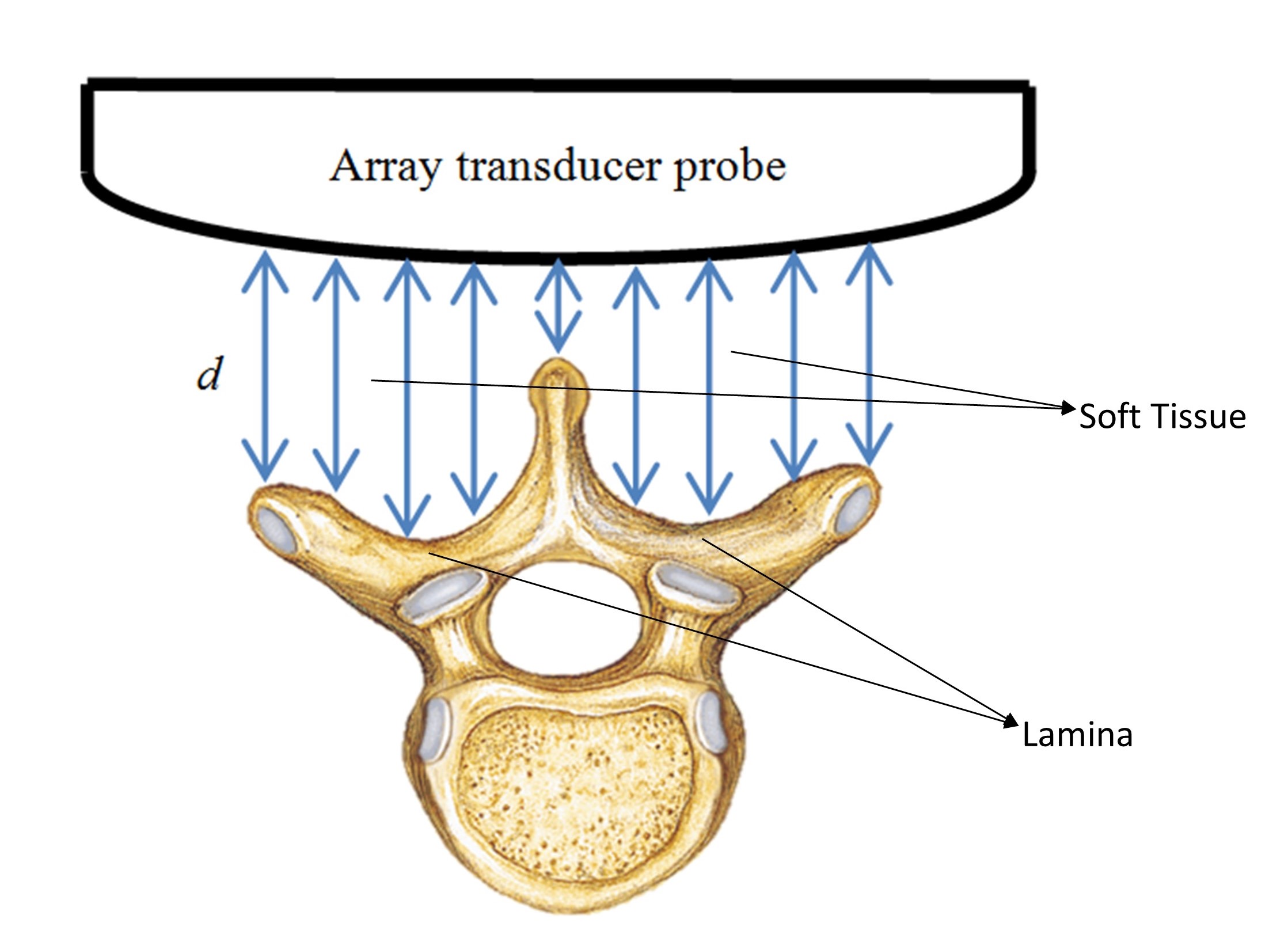}
	\caption{Ultrasound pulse reflects at the interface of soft tissue and lamina.}
\end{figure}
\(z\) is traveling distance of the ultrasound waves reflected from the cortex/soft-tissue interface, where in this circumstance \(z=2d\) and \(d\) is the thickness of the soft tissue layer. Fig.1 shows an ultrasound pulse travels though soft tissue and reflects at the surface of vertebral cortex. Considering the frequency amplitude of the wave,
Eq.(\ref{eq1}) is then converted by taking logarithm as 
\begin{equation}
	\ln \left|A\left(f,z=2 d\right)\right|=\ln R+\ln |S(f)|-2 \alpha d
	\label{eq2}
\end{equation}
Define the amplitude \(A_{d B}=20 \log _{10}(A)=8.686 \ln (\mathrm{A})\), then Eq.(\ref{eq2}) can be rewritten as
\begin{equation}
	A_{d B}(f)=R_{d B}+S_{d B}(f)-2 \alpha_{d B} d
	\label{eq3}
\end{equation}
We define the maximum amplitude of the frequency spectrum as the frequency amplitude index (FAI), and it indicates the energy peak at the main frequency of an ultrasound reflection signal. Given the constant source pulse \(S(f)\), the 
FAI is only influenced by the reflection coefficient \((R_{dB})\), the attenuation coefficient \((\alpha_{dB})\) and the thickness of soft tissue \((d)\). In this case, 
the reflection coefficient can be calculated as 
\begin{equation}
	R = \frac{\left( Z_2 - Z_1 \right)^2}{\left( Z_2 + Z_1 \right)^2}
	\label{eq4}
\end{equation}
where \(Z_1\) and \(Z_2\) are the acoustic impedance of soft tissue and vertebral cortex respectively. 
According to \citet{goss_comprehensive_1978,goss_compilation_1980}, the range of acoustic impedances were \(3.75-7.38*10^6kg/(m^2\cdot s)\) for bone tissue and \(1.65-1.74*10^6kg/(m^2\cdot s)\) for muscles, and it resulted in the range of reflection coefficients \((R_{dB})\) was -8 \textasciitilde -18dB. 
On the other hand, the attenuation coefficient \((\alpha_{dB})\) for soft tissue at 2MHz was about 1dB/cm \citep{kremkau_diagnostic_2002}; the difference of the soft tissue thickness among subjects was usually only a few millimeters \citep{vermess_normal_2012,zheng_simultaneous_2013}, and it demonstrated 
that the attenuation caused by soft tissue was mostly less than 1dB for pediatrics. Therefore the FAI is mainly influenced by the reflection coefficients \((R)\) of the interface between vertebral cortex and soft tissue, 
and moreover substantially affected by the bone parameters including acoustic impedance, density and velocity.

\subsection*{Bone specimens}      
Eight bovine femur specimens were obtained from a local butcher shop and applied for the in vitro study to validate the FAI method. The specimens were cut to 30mm*20mm*6mm blocks. The nitric acids solution with concentration of 1\%, 2\%, 4\% and 7\% was applied to 
decalcify the specimens. Eight specimens were randomly divided into two groups (n = 4 in each), and the specimens in each group were decalcified for 6 h and 12 h, respectively. The density of the specimens was calculated from the mass and volume. 
The mass of specimens was measured by electronic balance, and the volume was defined as product of length, width and height. The velocities of the specimens were measured using the ultrasound 
transmission method \citep{zheng_simultaneous_2013} with one pair of 2.25MHz immersion transducers (Doppler, Guangzhou, China). The velocity of water was also measured as the reference. The physical and acoustical parameters of all specimens were 
listed in Table 1, and the reflection coefficients were calculated by Eq.(\ref{eq4}).
\begin{table}
	\caption{Physical and acoustical parameters of the decalcified bovine bone specimens.}
	\resizebox{\textwidth}{!}{
	\centering
	\begin{tabular}{@{}ccccccc@{}}
		\toprule
		Number &
			\begin{tabular}[c]{@{}c@{}}Density\\ (\(g/cm^3\))\end{tabular} &
			\begin{tabular}[c]{@{}c@{}}Velocity\\ (\(m/s\))\end{tabular} &
			\begin{tabular}[c]{@{}c@{}}Acoustic\\ Impedance\\ (Mrayl)\end{tabular} &
			\begin{tabular}[c]{@{}c@{}}Reflection\\ Coefficient*\end{tabular} &
			\begin{tabular}[c]{@{}c@{}}Nitric Acid\\ Concentration\end{tabular} &
			\begin{tabular}[c]{@{}c@{}}Decalcification\\ Duration(h)\end{tabular} \\ \midrule
		1 & 1.927 & 3285 & 6.331 & 0.391 & 1\% & 6  \\
		2 & 1.882 & 2710 & 5.101 & 0.308 & 1\% & 12 \\
		3 & 1.905 & 2966 & 5.651 & 0.348 & 2\% & 6  \\
		4 & 1.847 & 2637 & 4.870 & 0.290 & 2\% & 12 \\
		5 & 1.668 & 2295 & 3.829 & 0.201 & 4\% & 12 \\
		6 & 1.845 & 2848 & 5.254 & 0.320 & 4\% & 6  \\
		7 & 1.760 & 2717 & 4.782 & 0.283 & 7\% & 6  \\
		8 & 1.473 & 2211 & 3.258 & 0.145 & 7\% & 12 \\ \bottomrule
	\end{tabular}
	}
*Calculated between the specimen and water (Density: 1.00\(g/cm^3\), Velocity: 1459\(m/s\)) using Eq.(\ref{eq4})
\end{table}
\subsection*{Subjects}
For the in vivo measurement, one hundred and one subjects (F:86, M:15, age:14.2±2.0) were recruited from the local scoliosis clinic. The inclusion criteria were: 1) diagnosed with idiopathic scoliosis, 2) 10-18 years old, 
3) no surgical treatment prior to the study day, 4) with at least one clinical record acquired during 12 months before or after the study day, and 5) with original RF ultrasound data.  Ethics approval was granted from the 
local health ethics board. All subjects participating in the study signed a written consent form. The information of Risser grade, age, body mass index (BMI), Cobb angles and curve progression status for the subjects at the study day were exported from the local clinical records. The curve was 
determined as progressed if the increase between the Cobb angles obtained from two successive clinical visits was \(\ge 6°\). One of these two visits was at the study day, and the other occurred within one year prior to or later than 
the study day.  

\subsection*{Data Acquisition and Analysis}
For the in vitro study, the specimens were immersed in the water tank, and the reflected RF data from the specimen surface were acquired using the Vantage ultrasound system equipped with C5-2v transducer (Verasonics Inc., 
Redmond, WA, USA). Each RF data contained 128 channel records, and 20 channels in the middle were selected to calculate FAI. For each record, Hamming window was used to 
select main pulse of the echo. For each frame, the FAI value was calculated on the average spectrum of the selected echo records.

To acquire the reflection signals from scoliotic subjects, the standing ultrasound scan was performed on the subject's back as described in our previous study \citep{zheng_intra-_2015} using the SonixTABLET system (Analogic Ultrasound – BK Medical,
 Peabody, MA, USA). Three adjacent B-frame images at the lumbar vertebra (L5) were selected based on the clarity of the reflection from vertebral surface, and the original RF data of the corresponding frames were then exported. 

\begin{figure}
	\centering
	\includegraphics[scale=0.6]{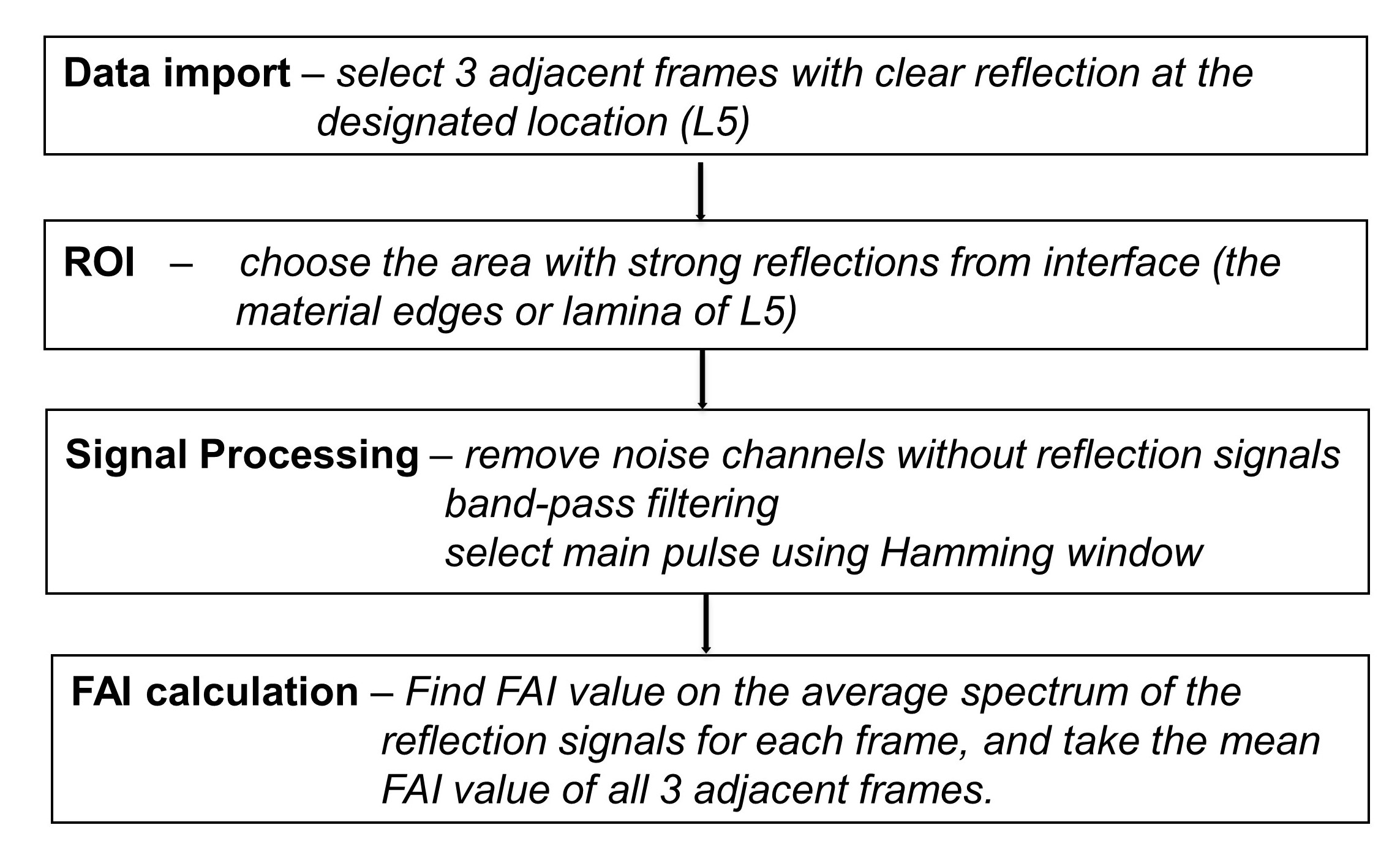}
	\caption{The flow chart of FAI calculation.}
\end{figure}
 The procedure as shown in Fig.2 was applied to calculate the FAI of in vivo ultrasound reflection signals from clinical subjects. An in-house program was developed to load and process the RF data and export FAI values for the 
 selected frame. Fig.3 showed the program interface and a typical RF frame data collected from a scoliotic subject, and the red color illustrated echoes from the different anatomic structures such as soft tissues and vertebra.
 Each RF frame data were consisted of 256 signal records, however only the area indicating strong reflections from lamina (flat surface of vertebra) was selected as the region of interest (ROI) and contained approximately 
 54-88 records (black box in Fig.3) for one frame. To minimize the random effect, the average was taken among the spectrums of all the signal records in ROI 
 for one frame, and the FAI value was calculated on the average spectrum using Eq.\ref{eq3} for the frame. Then the mean of FAI values calculated over three selected adjacent frames were used to demonstrate the FAI for the subject. Therefore, the FAI value in this study implied the average energy 
 of the ultrasound signals reflected from the surface of lumbar vertebra, which should be correlated to the reflection coefficient of bone/tissue interface and moreover related to the quality of bone.  
\begin{figure}[h]
	\centering
	\includegraphics[scale=0.7]{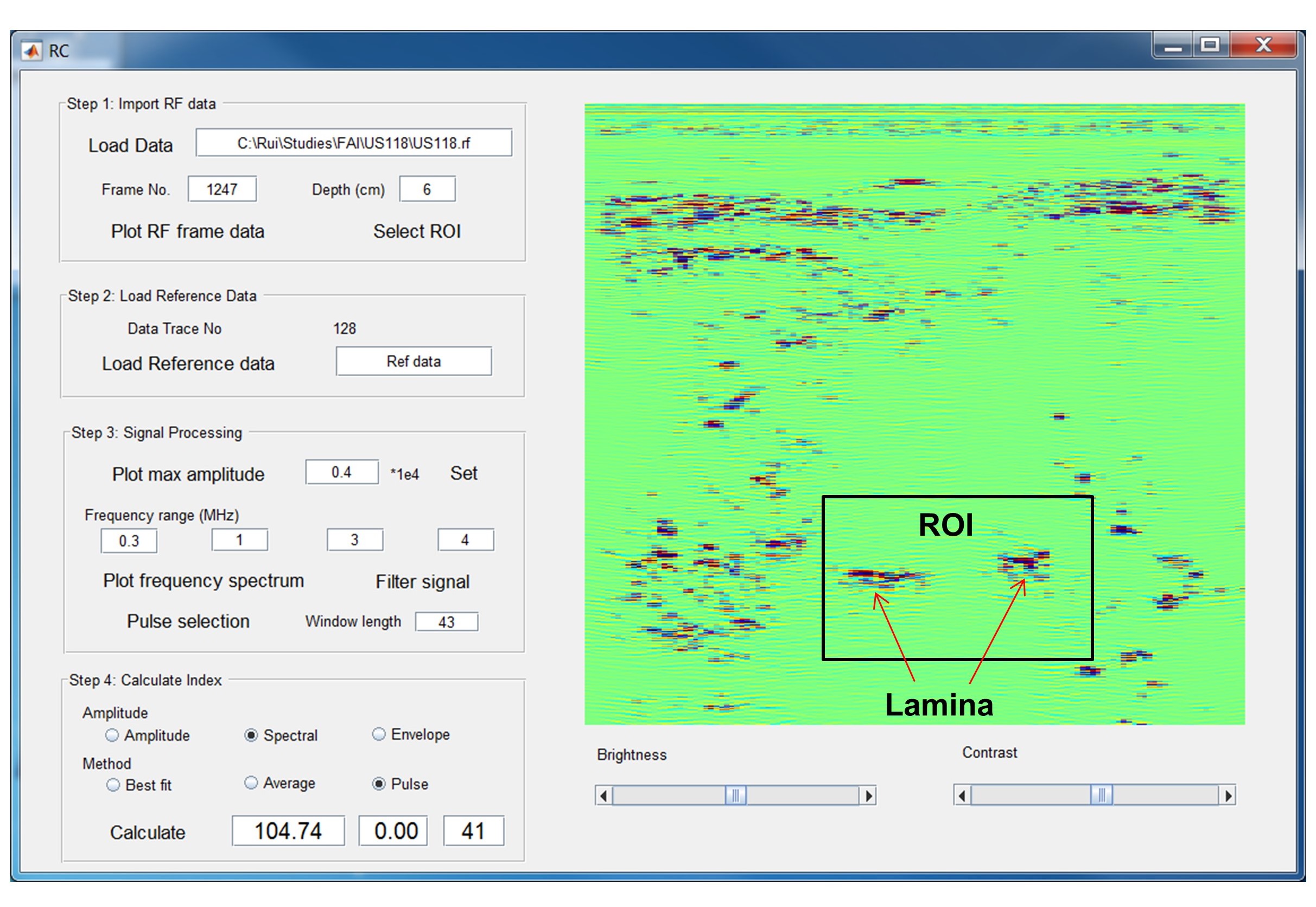}
	\caption{The in-house program interface for FAI calculation.}
\end{figure}

All FAI values were exported in dB. The standard deviations (SD) were calculated among three selected adjacent frames at L5 for each subject to examine the error range of the FAI estimation. 
One rater with more than 10-year experience on ultrasound research measured all the data twice with one week apart (2 trials), and the mean absolute difference (MAD),
standard deviation, the intra-class correlation coefficients using a 2-way random model and absolute agreement with a 95\% confidence interval (ICC[2,1])  and the standard errors of measurement (SEM) were 
calculated to assess the repeatability and reliability of the method. The FAI from the second trial was used  to analyze the influences  referring to gender, BMI, maximum Cobb angles and the curve progression status.  
\section*{Results}
\label{Results}
Fig.4 illustrated the FAI values referring to the bone parameters including density, velocity, acoustic impedance and reflection coefficient of bovine specimens, respectively. The correlation coefficients \((R^2)\) 
were all higher than 0.70, and the FAI showed the highest correlation with the reflection coefficient \((R^2=0.824)\).
\begin{figure}
	\centering
	\includegraphics[scale=0.8]{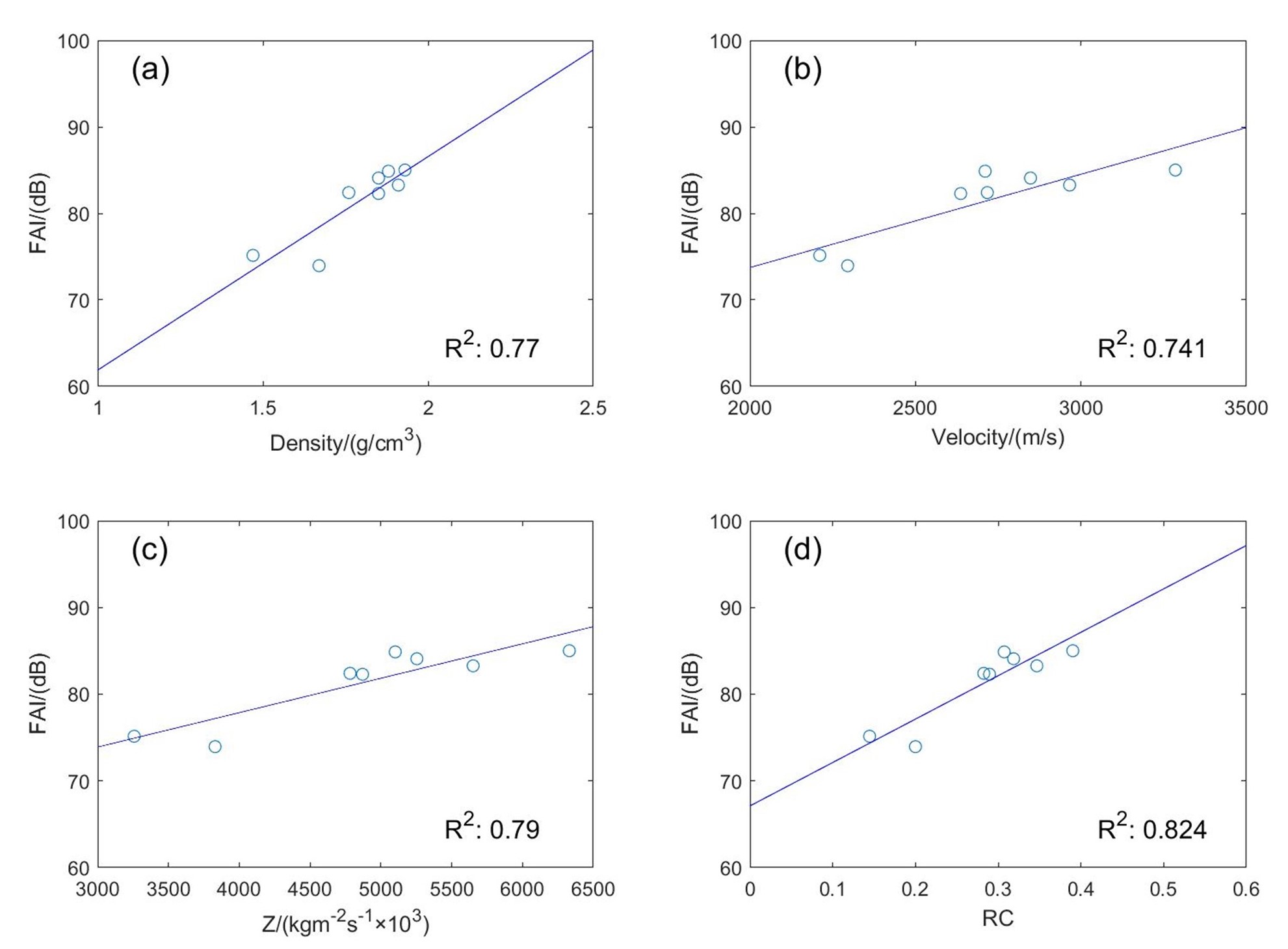}
	\caption{The FAI value in reference to (a)density, (b)velocity, (c)acoustic impedance and (d)reflection coefficient of bone specimens.}
\end{figure}

\begin{table}
	\caption{Comparison for the FAI measurement between two trials from the same rater.}
	\centering
	\begin{tabular}{@{}cccccc@{}}
		\toprule
		\begin{tabular}[c]{@{}c@{}}Trial \end{tabular} &
		  \begin{tabular}[c]{@{}c@{}}Mean (SD)\\(dB)\end{tabular} &
		  \begin{tabular}[c]{@{}c@{}}Range\\(dB)\end{tabular} &
		  \begin{tabular}[c]{@{}c@{}}MAD (SD)\\(dB)\end{tabular} &
		  ICC{[}2,1{]} &
		  \begin{tabular}[c]{@{}c@{}}SEM\\(dB)\end{tabular} \\ \midrule
		1 &
		  101.4 (2.2) &
		  95.2-105.6 &
		  \multirow{2}{*}{0.13 (0.11)} &
		  \multirow{2}{*}{0.997} &
		  \multirow{2}{*}{0.12} \\ \cmidrule(r){1-3}
		2 &
		  101.3 (2.2) &
		  95.2-105.6 &
		   &
		   &
		   \\ \bottomrule
		\end{tabular}
\end{table}
Table 2 demonstrated the intra-rater difference between the two trials on all 101 scoliotic subjects. The results showed the FAI values from the two measures were all in the same range, and the relative mean difference  between the two trials were only 0.13\%. 
The measures showed very reliable results with ICC of 0.997 \citep{currier_elements_1990}. For each subject, the FAIs were also calculated for the three adjacent frames, and the average SD among adjacent frames over all 101 subjects was 0.19 and 0.18 dB 
for two trials respectively. The measurement differences regarding to different frames and trials were all less than 0.2dB, which indicated the mean estimation error of FAI measurement would be smaller than 0.2dB as well.

The mean FAI was 101.2±2.2dB for 86 female subjects and 102.2±2.2dB for 15 male subjects. The mean value showed 1.0dB difference between different genders, which was 5 times larger than the measurement error for individual 
subject (0.2dB). Therefore only the FAI of female subjects was used for further analysis.

\begin{figure}[ht]
	\centering
	\includegraphics[scale=0.5]{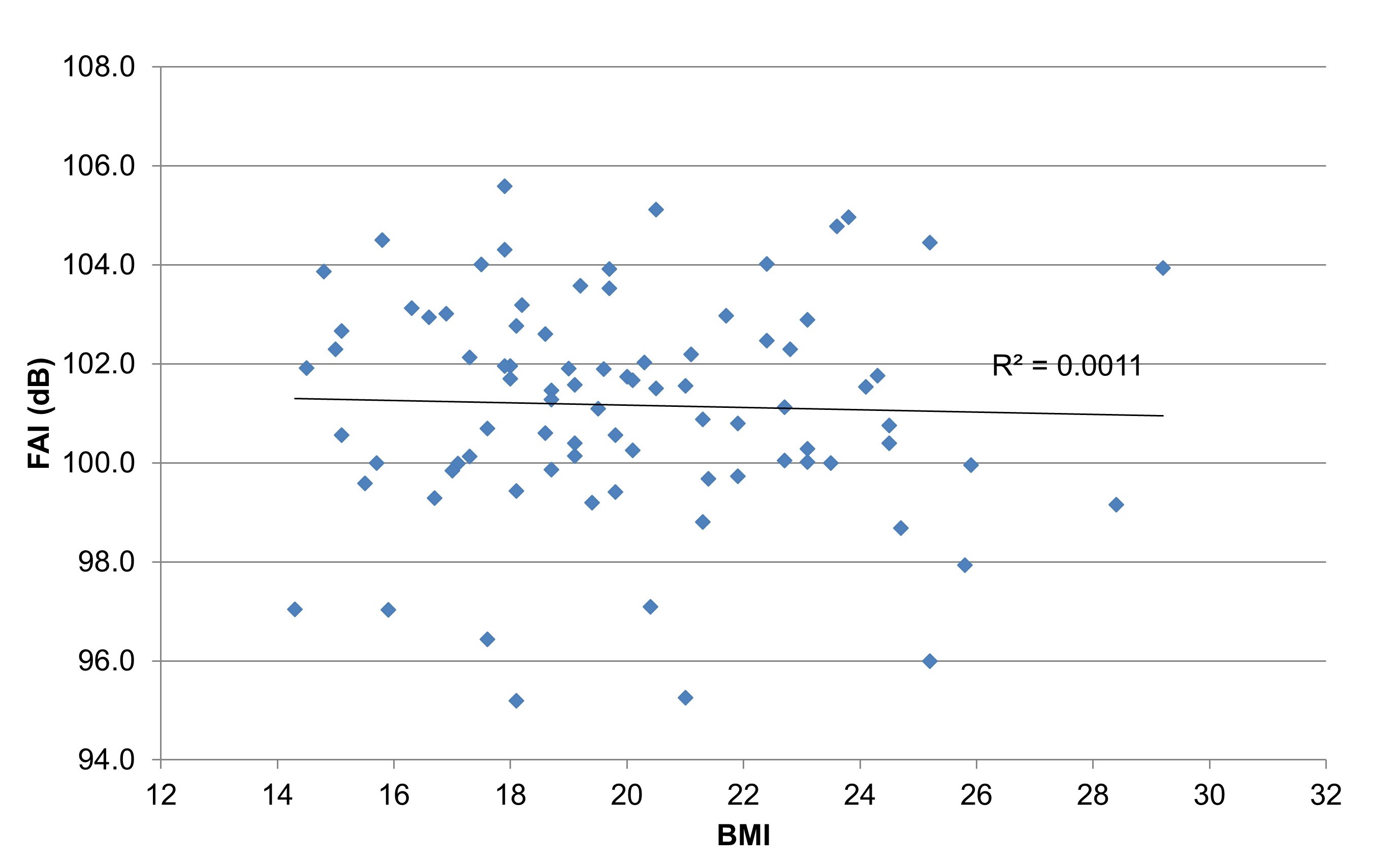}
	\caption{The FAI value in reference to subjects’ BMI.}
\end{figure}

\begin{figure}[ht]
	\centering
	\includegraphics[scale=0.5]{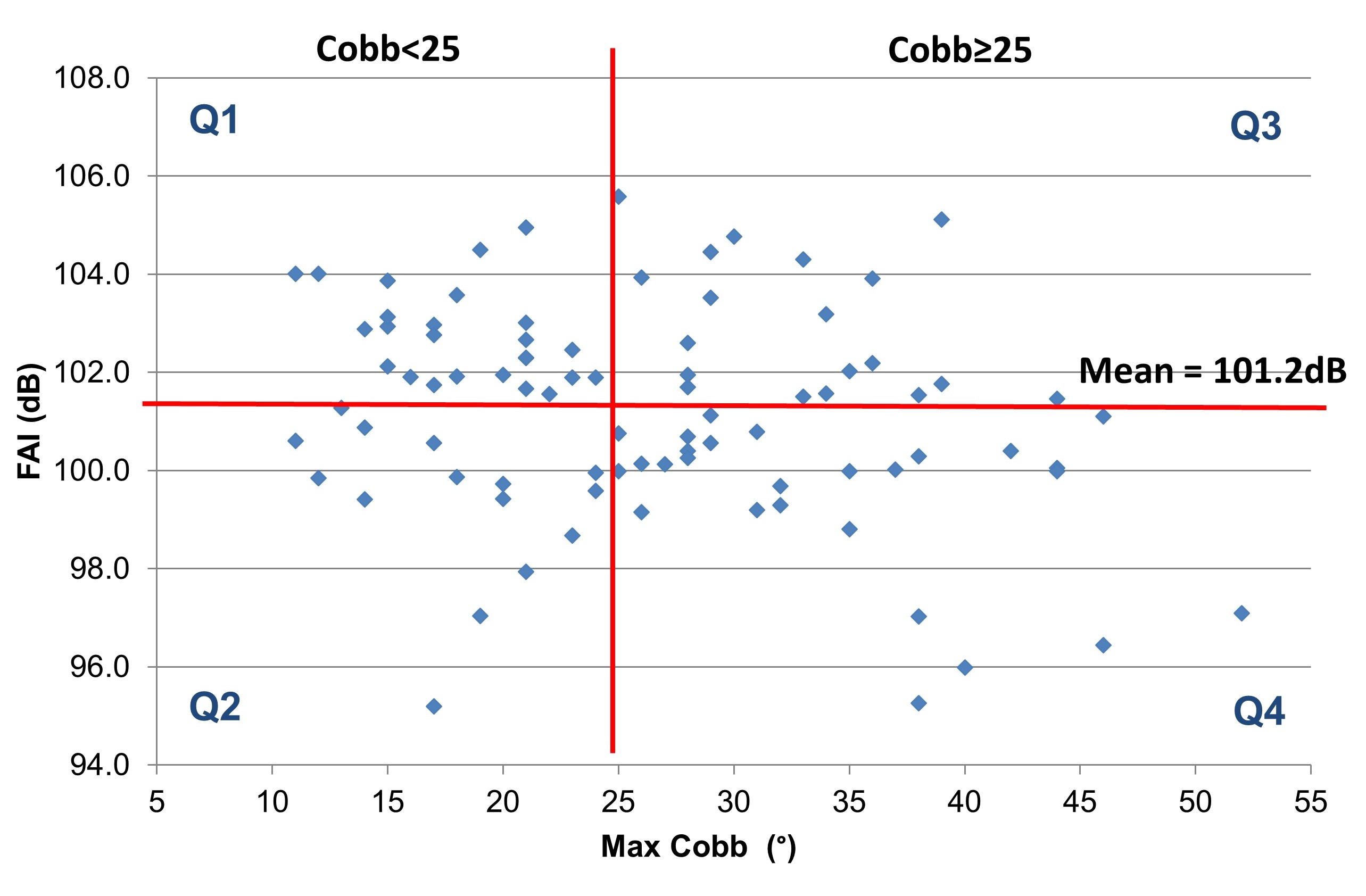}
	\caption{The FAI value in reference to subjects’ max Cobb angle.}
\end{figure}

Fig.5 and Fig.6 illustrated the FAI values in reference to BMI and maximum Cobb angles for all 86 female subjects. The distribution revealed no apparent difference of FAI values for BMI in the range of 14.3-29.2 and Cobb angles in the range of 11°-52°. 
In Fig.6, all female subjects were divided into 4 zones (Q1-Q4) according to the categories of mild curve (Cobb\(<\)25°) vs non-mild curves (Cobb\(\geq\)25°) and the categories of high FAI value (FAI\(\geq\)101.2dB) vs low FAI value (FAI\(<\)101.2dB), 
where 101.2dB was the mean FAI from all female subjects. Table 3 summarized the subject and curve information in four zones. There did not show apparent differences for Risser grade, age and BMI among the subjects in Q1-Q4. 
The mean and standard deviation of FAI were \(101.5\pm 2.0\)dB for mild curve group (Q1+Q2) and \(100.9\pm 2.3\)dB for non-mild curve group (Q3+Q4), respectively. 

\begin{figure}[ht]
	\centering
	\includegraphics[scale=0.5]{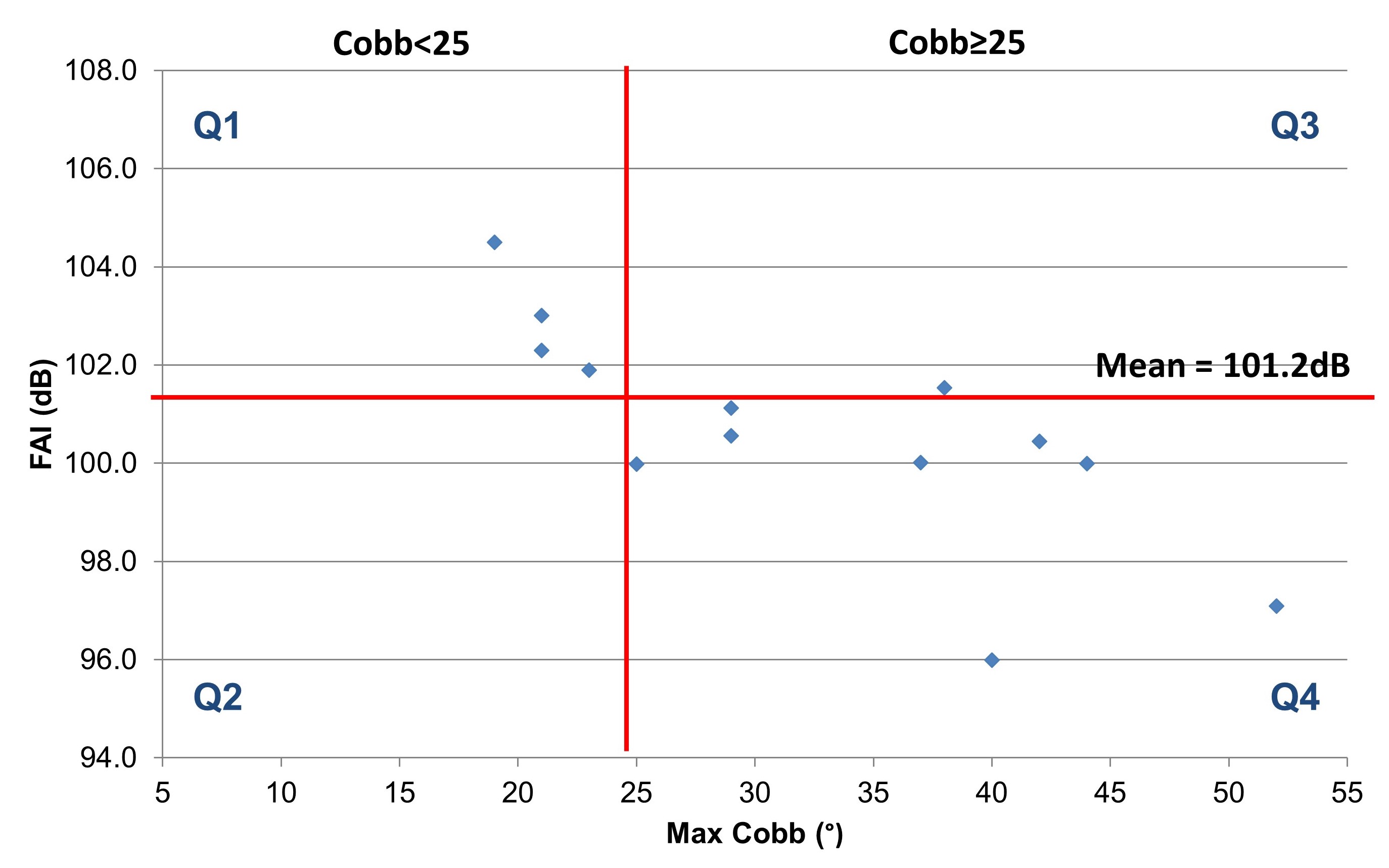}
	\caption{The FAI distribution of progression cases.}
\end{figure}
Table 4 listed progression status of all female subjects regarding to different zones. In the non-mild curve zones (Q3+Q4), progression cases showed higher prevalence than the mild curve zones (Q1+Q2). However, 
there were none progression cases in Q2 and only one in Q3, and the prevalence of curve progression in these two zones were much lower than Q1 \& Q4. Fig.7 illustrated 
the distribution of all progression cases referring to the maximum Cobb angle and FAI value.
\section*{Discussion}
\label{Discuss}
In the in vitro experiment, the FAI showed strong correlations with the properties of bone tissue. As indicated in Fig.4, the FAI were highly correlated with density, velocity, acoustic impedance and reflection coefficient 
\((R^2>0.70)\). Since the FAI implied the energy of echoes from the interface between vertebra and soft tissue, the reflection coefficient showed the highest correlation among all parameters. On the other hand, the \(R^2\) between FAI and 
acoustic impedance was 0.79, which suggested that the FAI could be regarded as a parameter to indicate bone quality. It was worth mentioned that the decalcification solution used in the experiment corroded some bone 
specimens. As the result, the surface of bone specimens became quite rough in several cases, which could disperse the energy of the received ultrasound echo and cause the large uncertainty of FAI calculation.

Based on the statistical analysis, the FAI method demonstrated good reliability and repeatability for the different trials and frames. This consistent measurement results were partly due to the simplified processing 
over the frequency domain. The procedure chose the main pulse of the echo and calculated the FAI value from the in the average frequency spectrum of echo signals collected in ROI, therefore the results represented the integrated energy of the reflection signals 
along the vertebral surface. The energy integration during signal analysis could reduce the variance and influence from the acquisition parameters, for example the selection of amplitude peak in time domain. Therefore, comparing 
to the amplitude analysis in time domain, the FAI presented more consistent and robust results with relative measurement difference between trials less than 1\%.

As illustrated in Fig.5, the FAI values were evenly distributed over the whole range of BMI (14.3-29.2), which indicated that the soft tissue on patient’s back had very limited influence on the measurement of FAI. It also implied  that the FAI value was mainly 
influenced by the reflection coefficient of vertebral surface and was a good indicator of bone quality.

Even though the mean FAI values of mild and non-mild curve groups were slightly different, 101.5dB and 100.9dB respectively, there was no apparently correlation between FAI and max Cobb angle \((R^2 = 0.0741)\) as show in Fig.6, 
which suggested that FAI method might not be suitable to directly estimate the severity of scoliosis.

In Table 3, we summarized the subjects’ information. There were no apparent differences for Risser grade, age and BMI, which illustrated the similarity of the patient groups from different zones. However as indicated in Fig.7, 
the FAI values of the progression cases were all located in certain zones. Especially for the non-mild curve group (Q3+Q4), the FAI of progression cases were all in the range lower than the mean value (Q4). It indicated that the subjects with lower FAI values had shown higher risk of progression for the group of non-mild curve.
A possible explanation was that the bone quality of spine could change and thus affect the reflection coefficient of bone surface during curve progression \citep{salzmann_skin_2019}, therefore resulted in the different performance of  
FAI values. Based on the statistics from Table 4, for those non-mild cases, the subjects with under-average FAI values exhibited higher probability to progress than the subjects with above-average FAI values (30\% vs 5\%). 
It illustrated that the FAI method had the potential to monitor the progression for a scoliotic patient. 

\begin{table}
	\caption{The summary of subject and curve information for Zone Q1-Q4.}
	\centering
	\resizebox{\textwidth}{!}{%
	\begin{tabular}{@{}cccccccccccc@{}}
		\toprule
		\multirow{2}{*}{Zone} &
		  \multirow{2}{*}{\begin{tabular}[c]{@{}c@{}}Number of\\ Subjects\end{tabular}} &
		  \multirow{2}{*}{Risser} &
		  \multicolumn{3}{c}{Cobb Angle (°)} &
		  \multicolumn{3}{c}{Age} &
		  \multicolumn{3}{c}{BMI} \\ \cmidrule(l){4-12} 
				&    &     & Mean & SD  & Range & Mean & SD  & Range     & Mean & SD  & Range     \\ \midrule
		Overall & 86 & 0-5 & 26.5 & 9.6 & 11-52 & 14.1 & 2.0 & 10.0-17.6 & 19.9 & 3.2 & 14.3-29.2 \\
		Q1      & 26 & 0-5 & 18.1 & 3.6 & 11-24 & 14.1 & 1.9 & 10.4-17.4 & 18.7 & 2.8 & 14.5-23.8 \\
		Q2      & 14 & 0-5 & 18.1 & 4.2 & 11-24 & 13.6 & 1.8 & 10.3-15.6 & 19.6 & 3.8 & 14.3-25.9 \\
		Q1+Q2   & 40 & 0-5 & 18.1 & 3.8 & 11-24 & 13.9 & 1.9 & 10.3-17.3 & 19.0 & 3.2 & 14.3-25.9 \\
		Q3      & 19 & 0-5 & 32.8 & 5.1 & 25-44 & 14.7 & 2.1 & 10.3-17.3 & 20.8 & 3.1 & 17.9-29.2 \\
		Q4      & 27 & 0-5 & 34.5 & 7.6 & 25-52 & 14.1 & 2.1 & 10.0-17.6 & 20.7 & 3.1 & 15.7-28.4 \\
		Q3+Q4   & 46 & 0-5 & 33.8 & 6.6 & 25-52 & 14.3 & 2.1 & 10.0-17.6 & 20.7 & 3.1 & 15.7-29.2 \\ \bottomrule
		\end{tabular}%
	}
\end{table}

\begin{table}
	\caption{The summary of progression cases in Zone Q1-Q4.}
	\centering
	\begin{tabular}{@{}cccc@{}}
		\toprule
		\multirow{2}{*}{Zone} & \multirow{2}{*}{\begin{tabular}[c]{@{}c@{}}Number of\\ Subjects\end{tabular}} & \multicolumn{2}{c}{Progression cases} \\ \cmidrule(l){3-4} 
				&    & Number & Prevalence* \\ \midrule
		Overall & 86 & 13     & 15\%                    \\
		Q1      & 26 & 4      & 15\%                    \\
		Q2      & 14 & 0      & 0\%                     \\
		Q1+Q2   & 40 & 4      & 10\%                    \\
		Q3      & 19 & 1      & 5\%                     \\
		Q4      & 27 & 8      & 30\%                    \\
		Q3+Q4   & 46 & 9      & 20\%                    \\ \bottomrule
	\end{tabular}

	*Calculated by dividing the number of progress cases to the number of subjects in the corresponding zones.
\end{table}
Finally, the FAI can be easily measured using exiting  medical ultrasound equipment for spine imaging. The data used to calculate FAI can be directly obtained during image acquisition without extra procedure. With good reliability and repeatability, the FAI can be used as a promising indicator to predict the curve progression of scoliotic patients in clinical application.  

\section*{Conclusions}
\label{Conclusions}
The paper proposed the frequency amplitude index (FAI) as a reliable parameter to assess the bone quality for scoliotic patients with the relative measurement error less than 1\%. The FAI method showed strong correlation \((R^2=0.824)\) with the reflection coefficient of bone tissue in vitro 
experiments and exhibited limited influence by BMI. Especially the subjects with lower FAI values in the non-mild curve group showed higher risk of curve progression (30\%). This preliminary study reported that the FAI method can provide 
a feasible and promising approach  to assess bone quality and monitor curve progression of the patients who have AIS.
        
\section*{Acknowledgements}
\label{Ack}
The authors are appreciated to the Glenrose Rehabilitation Hospital, Alberta Health Services, Canada, for clinical data acquisition in this study. This work was sponsored by Natural Science Foundation of Shanghai (Grant No.19ZR1433800). 

\section*{Conflict of interest}
The authors declare that they have no conflicts of interest to this work.





\pagebreak

\bibliographystyle{UMB-elsarticle-harvbib}
\bibliography{reference}

\end{document}